\documentclass[submission]{eptcs}
\providecommand{\event}{TARK 2015} 

\usepackage{amsfonts}
\usepackage{mathrsfs}
\usepackage{amssymb}
\usepackage{latexsym}
\usepackage{graphicx}
\usepackage{enumerate}
\usepackage{mathtools}
\usepackage{color}
\usepackage{float}
\usepackage{array}

\newcommand{\shortv}[1]{#1}
\newcommand{\fullv}[1]{}

\usepackage{tabularx}

\mathtoolsset{centercolon}

\fullv{\usepackage{chicago}}

\newtheorem{THEOREM}{Theorem}[section]
\newenvironment{theorem}{\begin{THEOREM} \hspace{-.85em} {\bf :} }%
                        {\end{THEOREM}}
\newtheorem{LEMMA}[THEOREM]{Lemma}
\newenvironment{lemma}{\begin{LEMMA} \hspace{-.85em} {\bf :} }%
                      {\end{LEMMA}}
\newtheorem{COROLLARY}[THEOREM]{Corollary}
\newenvironment{corollary}{\begin{COROLLARY} \hspace{-.85em} {\bf :} }%
                          {\end{COROLLARY}}
\newtheorem{PROPOSITION}[THEOREM]{Proposition}
\newenvironment{proposition}{\begin{PROPOSITION} \hspace{-.85em} {\bf :} }%
                            {\end{PROPOSITION}}
\newtheorem{DEFINITION}[THEOREM]{Definition}
\newenvironment{definition}{\begin{DEFINITION} \hspace{-.85em} {\bf :} \rm}%
                            {\end{DEFINITION}}
\newtheorem{CLAIM}[THEOREM]{Claim}
\newenvironment{claim}{\begin{CLAIM} \hspace{-.85em} {\bf :} \rm}%
                            {\end{CLAIM}}
\newtheorem{EXAMPLE}[THEOREM]{Example}
\newenvironment{example}{\begin{EXAMPLE} \hspace{-.85em} {\bf :} \rm}%
                            {\end{EXAMPLE}}
\newtheorem{REMARK}[THEOREM]{Remark}
\newenvironment{remark}{\begin{REMARK} \hspace{-.85em} {\bf :} \rm}%
                            {\end{REMARK}}

\newcommand{\thm}{\begin{theorem}}
\newcommand{\lem}{\begin{lemma}}
\newcommand{\pro}{\begin{proposition}}
\newcommand{\dfn}{\begin{definition}}
\newcommand{\rem}{\begin{remark}}
\newcommand{\xam}{\begin{example}}
\newcommand{\cor}{\begin{corollary}}
\newcommand{\prf}{\noindent{\bf Proof:} }
\newcommand{\ethm}{\end{theorem}}
\newcommand{\elem}{\end{lemma}}
\newcommand{\epro}{\end{proposition}}
\newcommand{\edfn}{\bbox\end{definition}}
\newcommand{\erem}{\bbox\end{remark}}
\newcommand{\exam}{\bbox\end{example}}
\newcommand{\ecor}{\end{corollary}}
\newcommand{\eprf}{\bbox\vspace{0.1in}}
\newcommand{\beqn}{\begin{equation}}
\newcommand{\eeqn}{\end{equation}}

\newcommand{\bbox}{\vrule height7pt width4pt depth1pt}
\newcommand{\qed}{\eprf}
\newcommand{\clm}{\begin{claim}}
\newcommand{\eclm}{\end{claim}}

\renewcommand{\phi}{\varphi}

\newcommand{\B}{{\cal B}}

\newcommand{\I}{{\cal I}}

\renewcommand{\P}{{\cal P}}

\newcommand{\ol}{\setlength{\itemsep}{0pt}\begin{enumerate}}
\newcommand{\eol}{\end{enumerate}\setlength{\itemsep}{-\parsep}}
\newcommand{\ul}{\setlength{\itemsep}{0pt}\begin{itemize}}
\newcommand{\dl}{\setlength{\itemsep}{0pt}\begin{description}}
\newcommand{\edl}{\end{description}\setlength{\itemsep}{-\parsep}}
\newcommand{\eul}{\end{itemize}\setlength{\itemsep}{-\parsep}}

\newcommand{\commentout}[1]{}

\newcommand{\bi}{\begin{itemize}}
\newcommand{\ei}{\end{itemize}}
\newcommand{\be}{\begin{enumerate}}
\newcommand{\ee}{\end{enumerate}}

\usepackage{amssymb,tikz}

\newcommand{\defin}[1]{\textbf{#1}}

\newcommand{\val}[1]{[\![ #1 ]\!]}

\shortv{\newcommand{\citeyear}{\cite}}

\title{Bayesian Games with Intentions}
\shortv{\author{
Adam Bjorndahl
\institute{Carnegie Mellon University\\
Philosophy\\
Pittsburgh, PA 15213}
\email{abjorn@andrew.cmu.edu}
\and
Joseph Y. Halpern
\institute{Cornell University\\
Computer Science\\
Ithaca, NY 14853}
   \email{halpern@cs.cornell.edu}
\and
Rafael Pass
 \institute{Cornell University\\
Computer Science\\
Ithaca, NY 14853}
\email{rafael@cs.cornell.edu}
}}
\shortv{
\def\titlerunning{Bayesian Games With Intentions}
\def\authorrunning{A. Bjorndahl, J. Y. Halpern, \& R. Pass}
}

\begin{document}

\maketitle

\begin{abstract}
We show that standard Bayesian games cannot represent the full
spectrum of belief-dependent preferences. However, by introducing a
fundamental distinction between \emph{intended} and \emph{actual}
strategies, we remove this limitation. We define \emph{Bayesian games
  with intentions}, generalizing both Bayesian games and psychological
games \cite{GPS}, and prove that Nash equilibria in psychological
games correspond to a special class of equilibria as defined in our
setting. 
\end{abstract}

\section{Introduction}

\emph{Type spaces} were introduced by John Harsanyi \citeyear{Harsanyi} as a
formal mechanism for modeling games of incomplete information where
there is uncertainty about players' payoff functions.
Broadly speaking,
types are taken to encode payoff-relevant information, a
typical example being how each participant values the items in an
auction. An important feature of this formalism is that types
also encode beliefs about types. Thus, a type encodes not only a
player's beliefs about other players' payoff functions, but a whole
\textit{belief hierarchy}: a player's beliefs about other players'
beliefs, their beliefs about other players' beliefs about other
players' beliefs, and so on.

This latter point has been enthusiastically embraced by the epistemic
game theory community, where type spaces have been co-opted for the
analysis of games of \textit{complete} information. In this context,
types encode beliefs about the \emph{strategies} used by other
players in the game as well as their types. So again, types
encode a belief hierarchy, but one that describes players' beliefs
about other players' beliefs\ldots about the other players' types and
strategies. In this framework, one can determine whether a player is
rational given her type and strategy;
that is, whether her strategy is such that she is making a best
response to her beliefs, as encoded by her type.
Thus, rationality, common belief of rationality, and so on
can be defined 
as events in the space of (profiles of) strategy-type pairs.
This opens the way to epistemic analyses of solution
concepts, among other applications
\cite{DS15}.
In this setting, types do not encode payoff-relevant information; rather,
they are simply a tool for describing belief
hierarchies about other players' (types and) strategies.

By contrast, in a \emph{Bayesian game}, types are payoff-relevant objects
in that utility depends on them, though the payoff-relevant information
they are taken to encode often includes such things as characteristics of
the players (strength, work ethic, etc.), or more generally any relevant
facts that may not be common knowledge.
There is typically assumed to be a prior probability on types (indeed,
often a common prior), so a 
type can still be viewed as encoding
beliefs on other types in this setting (a type $t$ encodes the
probability obtained by conditioning the prior on $t$), and thus a
belief hierarchy. However, the only aspect of this belief hierarchy
that is typically used in Bayesian games is the first-order belief about
other players' types (but not beliefs about beliefs, and so on), which
is needed to defined a player's expected utility.
Nonetheless, it is possible to leverage the fact that types
encode beliefs to define Bayesian games in which players' preferences
depend to some extent on the \textit{beliefs}
of their opponents (see Example \ref{exa:emb}).
This observation is the point of departure for the present work.

The notion that a player's preferences might depend on the beliefs of
her opponents (or on her own beliefs) is not new.
\emph{Psychological games} \cite{BD09,GPS} model phenomena like
anger, surprise, and guilt by incorporating belief hierarchies
directly into the domain of the utility functions.
\emph{Language-based games} \cite{BHP13} model similar
belief-dependent preferences by defining utility over descriptions in
a given language (in particular, a language that can express the players'
beliefs). Types play no explicit role in either of these frameworks;
on the other hand, the discussion above suggests that they may be
naturally employed to accomplish many of the same modeling goals. 
Since Bayesian games and, more generally, type spaces have become
cornerstones of modern game theory, if the modeling and analysis of
psychological games could be carried out in this familiar framework,
it would unify these paradigms and thereby amplify both the insights
and the accessibility of the latter.
In this paper, we provide an extension of Bayesian games that allows
us to do just this.

There is an obvious obstruction to capturing general belief-dependent
preferences using types in the standard way: types in Bayesian games
encode beliefs about types, not about strategies.
This severely limits the extent to which preferences over types can
capture feelings like surprise or guilt, which are typically
expressed by reference to beliefs about strategies
(e.g., my opponent
is surprised if I do not play the strategy that she was expecting me
to play). It may seem that there is a simple
solution to this problem:
allow types to encode beliefs about strategies.  
But doing this leads to difficulties in the definition of
\emph{Bayesian Nash
equilibrium}, the standard solution concept in Bayesian games; this notion
depends on being able to freely associate strategies with types. In
Section \ref{sec:bay}, we give the relevant definitions and make this
issue precise. 

In Section \ref{sec:int}, we develop a modification of the standard
Bayesian setup where
each player is associated with \textit{two} strategies:
an \emph{intended} strategy that is determined
by her type (and thus can be the object of beliefs),
and an \emph{actual} strategy that is independent of her type (as in
standard Bayesian games). This gives us what we call \emph{Bayesian
games with intentions}. We define a solution concept for such games
where we require that, in equilibrium, the actual
and intended strategies are equal. As we show, under this requirement,
equilibria do not always exist.

In Section \ref{sec:psy}, we show that psychological games can be embedded
in our framework. Moreover,
we show that the notion of Nash equilibrium for psychological games defined
by Geanakoplos, Pearce, and Stachetti \citeyear{GPS} (hereafter GPS)
corresponds to
a special case of our own notion of equilibrium.
Thus, we realize all the advantages of psychological games in an arguably
simpler, better understood setting. We do not require complicated
beliefs hierarchies; these are implicitly encoded by types.


The advantages of distinguishing actual from intended strategies go
well beyond psychological games. As we show
\fullv{by example, this
distinction has a variety of applications. In Section \ref{sec:rdp}, we
focus on one such application, demonstrating that }%
\shortv{in the full paper, }%
intended strategies can
be fruitfully interpreted as \emph{reference points} in the style of
prospect theory \cite{KT79}.
One of the central insights of prospect theory is that the subjective
value of an outcome can depend, at least in part, on how that outcome
compares to some ``reference level''; for example, whether it is
viewed as a relative gain or loss. The intended/actual distinction
naturally implements the needed comparison between ``real'' and
``reference'' outcomes. 
\fullv{
K\H{o}szegi and Rabin \citeyear{KR06} (hereafter KR) have likewise
advanced a
model of reference-dependent preferences that
endogenizes reference points (by identifying them with players'
expectations); indeed, we take our cue from them in recognizing the
value of this kind of endogenization.
As a concrete illustration of our proposal,
we show that the KR framework
is subsumed by Bayesian games with intentions.}
Using this insight, we show that 
\emph{reference-dependent preferences}, as defined by
K\H{o}szegi and Rabin \citeyear{KR06}, can be captured using 
Bayesian games with intentions.

\section{Bayesian games} \label{sec:bay}

\subsection{Definition}

A \emph{Bayesian game} is a model of strategic interaction among players whose preferences can depend on factors beyond the strategies they choose to play. These factors are often taken to be characteristics of the players themselves, such as whether they are industrious or lazy, how strong they are, or how they value certain objects. Such characteristics can be relevant in a variety of contexts: a job interview, a fight, an auction, etc.

A \emph{type} of player $i$ is often construed as encoding precisely such characteristics. More generally, however, types can be viewed as encoding any kind of information about the world that might be payoff-relevant. For example, the resolution of a battle between two armies may depend not only on what maneuvers they each perform, but also on how large or well-trained they were to begin with, or the kind of terrain they engage on. Decision-making in such an environment therefore requires a representation of the players' uncertainty regarding these variables.

We now give a definition of Bayesian games that is somewhat more
general than the standard definition. This will make it easier for us to
develop the extension to Bayesian games with intentions. We
explain the differences after we give the definition.

Fix a set of \emph{players}, $N = \{1, \ldots, n\}$. A \defin{Bayesian game (over $N$)} is a tuple $\B = (\Omega, (\Sigma_{i}, T_{i}, \tau_{i}, p_{i},\\ u_{i})_{i \in N})$ where
\begin{itemize}
\item
$\Omega$ is the measurable space of \emph{states of nature};
\item
$\Sigma_{i}$ is the set of \emph{strategies available to player $i$};
\item
$T_{i}$ is the set of \emph{types of player $i$};
\item
$\tau_{i}: \Omega \to T_{i}$ is \emph{player $i$'s signal function};
\item
$p_{i}: T_{i} \to \Delta(\Omega)$ associates with each type $t_{i}$ of
  player $i$ a probability measure $p_{i}(t_{i})$ on $\Omega$
  with $p_{i}(t_{i})(\tau_{i}^{-1}(t_{i})) = 1$, representing
  \emph{type $t_{i}$ of player $i$'s beliefs} about the state of
  nature;%
\footnote{As usual, we denote by $\Delta(X)$ the set of probability
  measures on the measurable space $X$. To streamline the
  presentation,
we suppress measurability assumptions here and elsewhere in the paper.}
\item
$u_{i}: \Sigma \times \Omega \to \mathbb{R}$ is \emph{player $i$'s utility function}.%
\footnote{Given a collection $(X_{i})_{i \in
  N}$ indexed by $N$, we adopt the usual convention of denoting by $X$
the product $\prod_{i \in N} X_{i}$ and by $X_{-i}$ the product
$\prod_{j \neq i} X_{j}$.}
\end{itemize}

As we said above, this definition of a Bayesian game is more general
than what is presented in much (though not all) of the
literature. 
There are two main differences.
First, we take utility to be defined over strategies and
\textit{states of nature}, rather than over strategies and types
(cf.~\cite{OR94} for a similar definition). This captures the
intuition that what is really payoff-relevant is \textit{the way the
  world is}, and types simply capture the players' imperfect knowledge
of this. Since the type signal function profile $(\tau_1,
\ldots,\tau_n)$ associates with each world a type profile, utilities
can depend on players' types.
Of course, we can always restrict
attention to the special case where $\Omega = T$ and where $\tau_{i}:
T \to T_{i}$ is the $i$th projection function; this is called the
\emph{reduced form}, and it accords with a common conception of types
as encoding all payoff-relevant information aside from
strategy choices (cf.~\cite{ft:gt}). 

The second respect in which this definition is more general than is
standard is in the association of an \textit{arbitrary} probability
measure $p_{i}(t_{i})$ to each type $t_{i}$. It is typically assumed
instead that for each player $i$ there is some fixed probability
measure $\pi_{i} \in \Delta(\Omega)$ representing her ``prior
beliefs'' about the state of nature, and $p_{i}(t_{i})$ is obtained by
conditioning these prior beliefs on the ``private information''
$t_{i}$ (or, more precisely, on the event
$\tau_{i}^{-1}(t_{i})$).\footnote{To ensure this is well-defined, it
  is also typically assumed that none of player $i$'s types are null
  with respect to $\pi_{i}$; that is, for all $t_{i} \in T_{i}$,
  $\pi_{i}(\tau_{i}^{-1}(t_{i})) > 0$.} When $\pi_{1} = \pi_{2} =
\cdots = \pi_{n}$, we say that the players have a \emph{common prior};
this condition is also frequently assumed in the literature. We adopt
the more general setup because it accords with a standard
presentation of type spaces as employed for the epistemic analysis of
games of complete information
\cite{DS15}, thus making it easier for
us to relate our approach to epistemic game theory.

The requirement that
$p_{i}(t_{i})(\tau_{i}^{-1}(t_{i})) = 1$ amounts to assuming that each
player is sure of her own type (and hence, her beliefs); that is, in
each state $\omega \in 
\Omega$, each player $i$ knows that the true state is among those
where she is of type $t_{i} = \tau_{i}(\omega)$, which is exactly
the set $\tau_{i}^{-1}(t_{i})$.

\subsection{Examples}

It will be helpful to briefly consider two simple examples of Bayesian games, one standard and one a bit less so.

\begin{example}
{First consider a simplified auction scenario where each participant $i
\in N$ must submit a bid $\sigma_{i} \in \Sigma_{i} = \mathbb{R}^{+}$
for a given item. Types here are conceptualized as encoding
valuations of the item up for auction: for each $t_{i} \in
T_{i}$, let $v(t_{i}) \in \mathbb{R}^{+}$ represent how much player
$i$ thinks the item is worth, and define player $i$'s utility $u_{i}:
\Sigma \times T$ by 
$$
u_{i}(\sigma, t) = \left\{ \begin{array}{ll}
v(t_{i}) - \sigma_{i} & \textrm{if $\sigma_{i} = \displaystyle \max_{j \in N} \sigma_{j}$}\\
0 & \textrm{otherwise.}
\end{array} \right.
$$
Thus, a player's payoff is 0 if she does not submit the highest bid,
and otherwise is equal to her valuation of the item less her bid
(for simplicity, this model assumes that in the event of a tie, every
top-bidding player gets the item). Note that the state space here is
implicitly taken to be identical to the space $T$ of type profiles,
that is, the game is presented in reduced form.
A type $t_{i}$ therefore tells us not only how valuable 
player $i$ thinks the item is ($v(t_{i})$), but also what beliefs
$p_{i}(t_{i}) \in 
\Delta(T)$ player $i$ has about how the \textit{other} players value
the item (and what beliefs they have about \textit{their} opponents,
and so on). The condition that $p_{i}(t_{i})(\tau^{-1}(t_{i})) = 1$
then simply amounts to the assumption that each player is sure of
her own valuation (as well as her beliefs about other players'
types).
} \qed 
\end{example}

\begin{example} \label{exa:emb}
{Next we consider an example where the Bayesian framework is leveraged
to model a player whose preferences depend on the beliefs of her
opponent. Consider a game where the players are students in a class,
with player $1$ having just been called upon by the instructor to
answer a yes/no question. Assume for simplicity that $N = \{1,2\}$,
$\Sigma_{1} = \{\textsf{yes}, \textsf{no}, \textsf{pass}\}$, and
$\Sigma_{2} = \{*\}$ (where $*$ denotes a vacuous move, so only player
$1$ has a real decision to 
make). Let $\Omega = \{w_{y}, w_{n}, v_{y}, v_{n}\}$, where,
intuitively, states with the subscript $y$ are states where ``yes'' is
the correct answer, while states with the subscript $n$ are states
where ``no'' is the correct answer. Let $T_{1} = \{t_{1}, t_{1}'\}$,
$T_{2} = \{t_{2}, t_{2}', t_{2}''\}$, and define the signal functions
by 
$$
\begin{array}{c}
\tau_{1}(w_{y}) = \tau_{1}(w_{n}) = t_{1}, \ 
\tau_{1}(v_{y}) = \tau_{1}(v_{n}) = t_{1}', \mbox{ and }\\
\tau_{2}(w_{y}) = \tau_{2}(w_{n}) = t_{2} \textrm{ and }
\tau_{2}(v_{y}) = t_{2}' \textrm{ and } \tau_{2}(v_{n}) = t_{2}''.
\end{array}
$$ 
Finally, assume that all of the subjective probability measures arise
by conditioning a common prior $\pi \in \Delta(\Omega)$ on the type of
the player in question; assume further that $\pi$ is the uniform
distribution. It follows that in each state, player $1$ is unsure of
the correct answer. On the other hand, while in states $w_{y}$ and
$w_{n}$, player $2$ is also unsure of the correct answer, in states
$v_{y}$ and $v_{n}$, player $2$ knows the correct answer. Moreover, in
states $w_{y}$ and $w_{n}$, player $1$ is sure that player $2$ does not
know the correct answer, whereas in states $v_{y}$ and $v_{n}$, player
$1$ is sure that player $2$ \textit{does} know the correct answer
(despite not knowing it himself). We can therefore use this framework
to encode the following (quite plausible) preferences for player $1$:
guessing the answer is preferable to passing provided player $2$ does
not know the right answer, but passing is better than guessing
otherwise. Set 
$$u_{1}(\mathsf{yes}, w_{y}) = u_{1}(\mathsf{yes}, v_{y}) = u_{1}(\mathsf{no}, w_{n}) = u_{1}(\mathsf{no}, v_{n}) = 5,$$
representing a good payoff for answering correctly; set
$$u_{1}(\mathsf{pass}, x) = -2 \textrm{ for all $x \in \Omega$},$$
representing a small penalty for passing regardless of what the correct answer is; finally, set
$$
\begin{array}{c}
u_{1}(\mathsf{yes}, w_{n}) = u_{1}(\mathsf{no}, w_{y}) = -5 \mbox{ and }\\
u_{1}(\mathsf{yes}, v_{n}) = u_{1}(\mathsf{no}, v_{y}) = -15,\end{array}
$$
representing a penalty for getting the wrong answer that is
substantially worse in states where player $2$ knows the correct
answer.
}

{It is easy to check that if player $1$ considers $w_{y}$ and $w_{n}$
to be equally probable, then her expected utility for randomly
guessing the answer is $0$, which is strictly better than passing
(passing, of course, always yields an expected utility of $-2$). By
contrast, if player $1$ considers $v_{y}$ and $v_{n}$ to be equally
probable, then her expected utility for randomly guessing is $-5$,
which is strictly worse than passing. In short, player 1's decision
depends on what she believes about the beliefs of player 2.} \qed
\end{example}

Example \ref{exa:emb} captures what might be thought of as
\emph{embarrassment aversion}, which is a species of belief-dependent
preference: player $1$'s preferences depend on what player 2
believes. It is worth being explicit about the conditions that make
this possible:
\begin{itemize}
\item[C1.] States in $\Omega$ encode a certain piece of information $I$
  (in this case, whether the correct answer to the given question is
  ``yes'' or ``no'').
\item[C2.] Types encode beliefs about states.
\item[C3.] Utility depends on types.
\end{itemize}
From C1--C3, we can conclude that preferences can depend on what the
players believe about $I$.  

Not all kinds of belief-dependent preferences can be captured in the
Bayesian framework. Suppose, for example, that the goal of player $1$
is to surprise her opponent by playing an unexpected strategy. More
precisely, suppose that $\Sigma_{1} = \{\sigma_{1}, \sigma_{1}'\}$ and
we wish to define $u_{1}$ in such a way that player $1$ prefers to 
play $\sigma_{1}$ if and only if player $2$ believes he will play
$\sigma_{1}'$. In contrast to Example \ref{exa:emb}, this scenario 
cannot be represented with a Bayesian game for the following simple
reason: \textit{states do not encode strategies}. In other words,
condition C1 is not satisfied if we take $I$ to be player 1's
strategy. Therefore, types cannot encode such
beliefs about strategies, so utility cannot be defined in a way
that depends on such beliefs.  

This suggests an obvious generalization of the Bayesian setting,
namely, encoding strategies in states. Indeed, this is the idea we
explore in this paper; however, it is not quite as straightforward a
maneuver as it might appear, primarily due to its interaction with the
mechanics of \emph{Bayesian Nash equilibrium}. 

\subsection{Equilibrium} \label{sec:eqi}

Part of the value of Bayesian games lies in the fact that a
generalized notion of Nash equilibrium can be defined in this
framework, for which the following notion plays a crucial role: a
\emph{behaviour rule} for player $i$ is a function $\beta_{i}: T_{i}
\to \Sigma_{i}$. In Bayesian games, we talk about behaviour rule
profiles being in equilibrium, just as in normal-form games, we talk
about strategy profiles being in equilibrium.
Intuitively, $\beta_{i}(t_{i})$ represents the
strategy that type $t_{i}$ of player $i$ is playing, so a player's
strategy depends on her type. 

From a technical standpoint, behaviour rules are important because
they allow us to associate a payoff for each player with each
\textit{state}, rather than strategy-state pairs. Since types encode
beliefs about states, this yields a notion of expected utility for
each type. A Bayesian Nash equilibrium is then defined to be a profile
of behaviour rules such that each type is maximizing its own expected
utility. 

More precisely, observe that via the signal functions $\tau_{i}$, a
behaviour rule $\beta_{i}$ associates with each state $\omega$ the
strategy $\beta_{i}(\tau_{i}(\omega))$. Thus, a profile $\beta$ of
behaviour rules defines an \emph{induced utility function}
$u_{i}^{\beta}: \Omega \to \mathbb{R}$ as follows: 
$$u_{i}^{\beta}(\omega) = u_{i}((\beta_{j}(\tau_{j}(\omega)))_{j \in N},\omega).$$
The beliefs $p_{i}(t_{i})$ then define the \emph{expected utility} for
each type: let $E_{t_{i}}(\beta)$ denote the expected value of
$u_{i}^{\beta}$ with respect to $p_{i}(t_{i})$.
Denote by $B_{i}$ the set of all behaviour
rules for player $i$. A behaviour rule $\beta_{i}$ is a \defin{best
  response to $\beta_{-i}$}
if, for each $t_{i} \in T_{i}$,
$\beta_{i}$ maximizes $E_{t_{i}}$:
$$(\forall \beta_{i}' \in B_{i})(E_{t_{i}}(\beta_{i}, \beta_{-i}) \geq E_{t_{i}}(\beta_{i}', \beta_{-i})).$$
Finally, a \defin{Bayesian Nash equilibrium} of the Bayesian game $\B$
is a profile of behaviour rules $\beta$ such that, for each $i \in N$,
$\beta_{i}$ is a best response to $\beta_{-i}$. A (mixed) Bayesian
Nash equilibrium is guaranteed to exist when the strategy and types
spaces are finite (see \cite{Tian09} for a more general
characterization of when an equilibrium exists). 

\section{Intention} \label{sec:int}

\subsection{Definition}

Behaviour rules map types to strategies, but
the underlying model does not enforce any
relationship between types and strategies (or between states and
strategies). Thus, behaviour rules do not provide a mechanism
satisfying condition C1 with $I$ taken to be a player's
strategies, so they do not allow us to express 
preferences that depend on beliefs about strategies. In order to
express such preferences, we must associate strategies with states
in the model itself.
Note that once we do this, utility functions depend on strategies in
two ways.  Specifically, since $u_{i}$ is
defined on the cross product $\Sigma \times \Omega$, players'
preferences depend on strategies both directly (corresponding to the
strategy-profile component of $u_i$'s input) and as encoded in states
(the second component of $u_i$'s input). To keep track of this
distinction, we call these \emph{actual} and \emph{intended}
strategies, respectively. 

Formally, a \defin{Bayesian game with instantiated intentions} (BGII)
is a tuple $\I = (\Omega, (\Sigma_{i}, T_{i}, \tau_{i}, s_{i}, p_{i},\\
u_{i})_{i \in N})$, where $s_{i}: T_{i} \to \Sigma_{i}$ is \emph{player
$i$'s intention function} and the remaining components are defined
as in a Bayesian game.
(The reason for this terminological mouthful will become clear in
Section \ref{sec:eq2}, where we define \emph{Bayesian games with
intentions}.)
Each $s_{i}$ associates with each type $t_{i}$
of player $i$ an \emph{intended strategy} $s_{i}(t_{i})$. Intuitively,
we might think of $s_{i}(t_{i})$ as the strategy that a player of type
$t_{i}$ ``intends'' or ``is planning'' to play (though may ultimately
decide not to); alternatively, it might be conceptualized as the
``default'' strategy for that type; it might even be viewed as the
``stereotypical'' strategy employed by players of type $t_{i}$. 
The former interpretation may be appropriate in a situation where we 
want to think of self-control; for example, a player who
intends to exercise, but
actually does not. The latter interpretation may be appropriate if we
think about voting. Wealthy people in Connecticut typically vote
Republican, but a particular player $i$ who is wealthy and lives in
Connecticut (this information is encoded in her type) votes Democrat.

We associate intended strategies with types rather than directly with
states by analogy to behaviour rules, in keeping with the modeling
paradigm where the personal characteristics of a player---including
her beliefs, decisions, \textit{and intentions}---are entirely
captured by her type. Nonetheless, the composition $s_{i} \circ
\tau_{i}: \Omega \to \Sigma_{i}$ does associate strategies with states
and so satisfies condition C1 (again, with $I$ being a player's
strategy); thus, players can have beliefs about strategies. This, in
turn, allows us to 
define utility so as to capture preferences that depend on
beliefs about strategies. 

\subsection{Examples}

The presentation of a BGII is made clearer by introducing the following notation for the set of states where player $i$ intends to play $\sigma_{i}$:
$$\val{\sigma_{i}} = (s_{i} \circ \tau_{i})^{-1}(\sigma_{i}) = \{\omega \in \Omega \: : \: s_{i}(\tau_{i}(\omega)) = \sigma_{i}\}.$$

\begin{example} \label{exa:sp1}
{Consider a $2$-player game in which player $1$'s goal is to surprise
her opponent. We take player 2 to be surprised if his beliefs about
what player 1 intends to play are dramatically different from what
player 1 actually plays. For definiteness, we take ``dramatically
different'' to mean that his beliefs about player 1's intended
strategy ascribe probability 0 to player 1's actual strategy.
Thus, we define player 1's utility 
function as follows: 
$$
u_{1}(\sigma,\omega) = \left\{ \begin{array}{ll}
1 & \textrm{if $p_{2}(\tau_{2}(\omega))(\val{\sigma_{1}}) = 0$}\\
0 & \textrm{otherwise.}
\end{array} \right.
$$
(Recall that $p_{2}(\tau_{2}(\omega))$ is a measure
on states, which is why we apply it to
$\tau_{1}^{-1}(s^{-1}_{1}(\sigma_{1}))$, that is, the set of states
$\omega$ where player 1's intended strategy,
$s_{1}(\tau_{1}(\omega))$, is equal to $\sigma_{1}$.) \qed 
}
\end{example}

\begin{example} \label{exa:bra}
{Next we consider an example introduced by GPS \cite{GPS}
called \textit{the bravery 
  game}. This is a $2$-player scenario in which player $1$ has the
only real decision to make: he must choose whether to take a
\emph{bold} action or a \emph{timid} action, so $\Sigma_{1} =
\{\textsf{bold}, \textsf{timid}\}$ (and $\Sigma_{2} = \{*\}$). The
crux of the game is the psychological factor, described by GPS as
follows: player $1$ prefers ``to be timid rather than bold, unless he
thinks his friends expect him to be bold, in which case he prefers not
to disappoint them'' \cite{GPS}. It is also stipulated that player $2$
prefers player $1$ to be bold, and also prefers to think of him as
bold. Define $q: T \to [0,1]$ by 
$$q(t) = p(t_{2})(\val{\textsf{bold}}),$$
and $\tilde{q}: T \to [0,1]$ by
$$\tilde{q}(t) = E_{t_{1}}(q),$$
where $E_{t_i}(f)$ denotes the expected value of $f$ with
respect to the measure $p(t_i)$. We can then represent the players'
preferences in a reduced-form BGII as follows: 
$$
u_{1}(\sigma, t) = \left\{ \begin{array}{ll}
2 - \tilde{q}(t) & \textrm{if $\sigma_{1}(t_1) = \textsf{bold}$}\\
3(1 - \tilde{q}(t)) & \textrm{if $\sigma_{1}(t_1) = \textsf{timid}$,}
\end{array} \right.
$$
$$
u_{2}(\sigma, t) = \left\{ \begin{array}{ll}
2(1 + q(t)) & \textrm{if $\sigma_{1}(t_1) = \textsf{bold}$}\\
1 - q(t) & \textrm{if $\sigma_{1}(t_1) = \textsf{timid}$.}
\end{array} \right.
$$
This representation closely parallels that given in \cite{GPS}, in
which $q$ and $\tilde{q}$ are understood not as functions of types,
but (implicitly) as functions of belief
hierarchies.\footnote{Additionally, GPS give the value of $q$, 
not by the probability that player $2$ assigns to player $1$
  being bold, but by player $2$'s \textit{expectation} of the
  probability $p$ with which player $1$ decides to be bold. We forgo
  this subtlety for the time being.} But this makes no difference to
the preferences this game encodes. For example, it is easy to see that
player $2$ prefers player $1$ to be bold, and all the more so when $q$
is high---that is, all the more so when she believes with high
probability that he will be bold.\footnote{It is not quite clear why
  GPS define player $2$'s payoff in the event that player $1$ is timid
  to be $1-q(t)$ rather than $1+q(t)$. This latter value preserves the
  preferences described while avoiding the implication that,
  assuming that player $1$ will be timid, player $2$ also prefers
  to \textit{believe} that he will be timid---this stands in
  opposition to the stipulation that player $2$ prefers to think of
  her opponent as bold.} Similarly, one can check that player $1$
prefers to be timid provided that $\tilde{q}(t) < \frac{1}{2}$; in other
words, provided that his expectation of his opponent's degree of
belief in him being bold is sufficiently low. 
}

{Why not define player $1$'s preferences directly in terms of the
beliefs of his opponent, rather than his expectation of these beliefs?
GPS cannot do so because of a technical limitation of the framework as
developed in \cite{GPS}; specifically, that a player's utility can
depend only on \textit{her own} beliefs. 
Battigalli and Dufwenberg \citeyear{BD09} correct this
deficiency. BGIIs do not encounter such limitations in the first
place. In particular, 
it is easy enough to redefine player $1$'s utility as follows: 
$$
u'_{1}(\sigma, t) = \left\{ \begin{array}{ll}
2 - q(t) & \textrm{if $\sigma_{1}(t_1) = \textsf{bold}$}\\
3(1 - q(t)) & \textrm{if $\sigma_{1}(t_1) = \textsf{timid}$.}
\end{array} \right.
$$
In this case, we find that player $1$ prefers to be timid provided $q(t)
< \frac{1}{2}$, or in other words, provided that his opponent's degree
of belief in him being bold is sufficiently low. \qed 
}
\end{example}

Observe that in neither of the preceding examples did we provide a
concrete BGII, in that we did not explicitly define the type spaces,
the intention functions, and so on. Instead, we offered general
recipes for implementing certain belief-dependent preferences (e.g.,
to surprise, to live up to expectations, etc.)~in arbitrary
BGIIs. Particular choices of type spaces and intention functions do
play an important role in equilibrium analyses; however, as
illustrated by the preceding two examples, at the modeling stage they
need not be provided up front. 

\subsection{Equilibrium} \label{sec:eq2}

We now define a notion of equilibrium for this setting.  As a first
step towards this definition, given a
BGII $\I$, we say that a profile of behaviour
rules $\beta$ is an \defin{equilibrium of $\I$} provided: 
\begin{enumerate}[(1)]
\item
$\beta$ is a Bayesian Nash equilibrium of the underlying Bayesian game: that is, each $\beta_{i}$ is a best response to $\beta_{-i}$ in precisely the sense defined in Section \ref{sec:eqi};
\item
for each player $i \in N$, $\beta_{i} = s_{i}$.
\end{enumerate}
This definition, and in particular condition (2), embodies the
conception of equilibrium as a steady state of play where
each player has correct beliefs about her opponents (and
is best responding to those beliefs).
In a BGII,
beliefs about the strategies of one's opponents are beliefs about
intended strategies (although, in equilibrium, a player will also have
beliefs about actual strategies).
On the other hand, since behavior rules associate strategies with
types and players have beliefs over types,
behaviour rules also induce beliefs about
strategies; in our terminology, these are beliefs about actual
strategies.
Condition (2) implies that these two beliefs coincide in 
equilibrium; in equilibrium, each type of each player
actually plays the strategy she intended to play
(which is exactly the strategy her opponents expected her to play).

Does condition (2) collapse the distinction between intended and
actual strategies, thereby returning us to the classical setting? It
does not. First, in a standard Bayesian game we could not even write
down a model where players' preferences depended on beliefs about
strategies. In addition, although we demand that intended and
actual strategies coincide in equilibrium, this restriction \textit{does
  not apply to the evaluation of best responses}. Recall that
$\beta_{i}$ is a best response to $\beta_{-i}$ if and only if 
$$(\forall \beta_{i}' \in B_{i})(E_{t_{i}}(\beta_{i}, \beta_{-i}) \geq E_{t_{i}}(\beta_{i}', \beta_{-i})).$$
Crucially, $\beta_{i}'$ need not be equal to $s_{i}$. In other words,
for $\beta_{i}$ to count as a best response, it must be at least as
good as all other behaviour rules, including those that recommend
playing a strategy distinct from that specified by $s_i$.

\begin{example} \label{exa:sp2}
{Consider a $2$-player reduced-form BGII with $\Sigma_{1} = \{a,b\}$,
$\Sigma_{2} = \{*\}$, $T_{1} = \{x,x'\}$, and $T_{2} = \{y,y'\}$, and
where 
$$p_{1}(x)(\{y\}) = p_{1}(x')(\{y'\}) = p_{2}(y)(\{x'\}) = p_{2}(y')(\{x\}) = 1.$$
Let $u_{1}$ be defined as in Example \ref{exa:sp1}, encoding player $1$'s desire to surprise her opponent:
$$
u_{1}(\sigma_{1}, *, t) = \left\{ \begin{array}{ll}
1 & \textrm{if $p_{2}(t_{2})(\val{\sigma_{1}}) = 0$}\\
0 & \textrm{otherwise.}
\end{array} \right.
$$
Suppose that $s_{1}(x) = s_{1}(x') = a$. Then, of course,
$p_{2}(y)(\val{a}) = p_{2}(y')(\val{a}) = 1$, and likewise
$p_{2}(y)(\val{b})\\ = p_{2}(y')(\val{b}) = 0$. It follows immediately
that the expected utility of playing $a$ for either type of player $1$
is equal to $0$ (since player $1$ is sure that this will not surprise
her opponent), whereas the expected utility of playing $b$ for either
type of player $1$ is equal to $1$ (since, in this case, player $1$ is
sure that this \textit{will} surprise her opponent). In particular, if
$\beta_{1} = s_{1}$, then $\beta_{1}$ is not a best response. Thus,
this particular BGII admits no equilibrium. 

Now suppose that $s_{1}(x) = a$ and $s_{1}(x') = b$. This is, of
course, a different BGII from the one considered in the previous
paragraph, but it differs only in the specification of player $1$'s
intentions. Moreover, in this BGII it is
not hard to check that $\beta_{1} = s_{1}$ is a best response and
therefore constitutes an equilibrium: type $x$ is sure that player $2$
is of type $y$; therefore, type $x$ is sure that player $2$ is sure
that player $1$ is of type $x'$, and so is playing $b$; thus, $a$ is a
best response for $x$, since $x$ is sure that it will surprise her
opponent; a similar argument shows that $b$ is a best response for
$x'$. \qed 
}
\end{example}

Example \ref{exa:sp2} demonstrates that the notion of best response in
a BGII---and therefore the notion of equilibrium---can be sensitive to
states of play where players are \textit{not} playing their intended
strategies. But it also illustrates the pivotal role of the intention
functions $s_{i}$ in determining the existence of an equilibrium. 
Indeed, condition (2) implies that if a given BGII $\I$ has an
equilibrium at all, it is unique and equal to $s$. This suggests that
BGIIs are not at the right ``resolution'' for equilibrium analysis,
since they come already equipped with a unique candidate for
equilibrium.
Thus, rather than restricting attention to a single BGII, where the
intention function is specified and hard-coded into the model, we
consider a more general model, 
where the intention function is not specified, but still 
affects the
utility. This is parallel to the role of strategies in standard games,
which are not hard-coded into the model, but of course the utility
function still depends on them.
Essentially,
we are moving the intention function from the model to the utility function.
As we shall see, our earlier examples of BGIIs
can be easily interpreted as models in this more general sense.

In order to make this precise, we must first
formally define utility functions that take as arguments intention functions.
More precisely, taking $\Sigma^{T} = \Sigma_{1}^{T_{1}} \times \cdots \times
\Sigma_{n}^{T_{n}}$ (so that $\Sigma^{T}$ is the set of intention
function profiles),
an \emph{explicit utility function}
is a map
$\tilde{u}_{i}: \Sigma \times \Omega \times \Sigma^T \to \mathbb{R};$
these are just like the utility functions in a BGII except they
explicitly take as input the associations between types and strategies
provided by intention functions.
A \defin{Bayesian game with intentions} (BGI)
  is a tuple $\tilde{\I} = (\Omega, (\Sigma_{i}, T_{i},
\tau_{i}, p_{i}, \tilde{u}_{i})_{i \in N})$, where the components are
defined just as they are in a Bayesian game, except that the
functions $\tilde{u}_{i}$ are explicit utility functions.
We emphasize that a BGI, unlike a BGII, does not include players'
intention functions among its components; instead, these functions
show up as arguments
in the (explicit) utility functions.

It is easy to see that all the examples of BGIIs that we have
considered so far can be naturally converted to BGIs. For example,
the utility function $u_1(\sigma,t)$ in Example~\ref{exa:sp1} becomes 
$\tilde{u}_i(\sigma,t,s)$. The definition of
$\tilde{u}_i(\sigma,t,s)$ looks identical to that of $u_1(\sigma,t)$;
the additional argument $s$ is needed to define $\val{\sigma_1}$.

A BGI induces a natural map from intention functions to BGIIs: given $\tilde{\I} = (\Omega, (\Sigma_{i}, T_{i}, \tau_{i}, p_{i}, \tilde{u}_{i})_{i \in N})$ and functions $s_{i}: T_{i} \to \Sigma_{i}$, let
$$\tilde{\I}(s_{1}, \ldots, s_{n}) = (\Omega, (\Sigma_{i}, T_{i}, \tau_{i}, s_{i}, p_{i}, u_{i})_{i \in N}),$$
where $u_{i}: \Sigma \times \Omega \to \mathbb{R}$ is defined by
$$u_{i}(\sigma, \omega) = \tilde{u}_{i}(\sigma, \omega, s_{1}, \ldots, s_{n}).$$
Clearly $\tilde{\I}(s_{1}, \ldots, s_{n})$ is a BGII; we call it an
\defin{instantiation of $\tilde{\I}$}. We then define an
\defin{equilibrium of $\tilde{\I}$} to be a profile of behaviour rules
$\beta$ that is an equilibrium of the corresponding instantiation
$\tilde{\I}(\beta)$. Here we make implicit use of the fact
that both behaviour rules and intention functions are 
functions from types to strategies.
Indeed, the profile $\beta$ plays two roles: first, it
is used to determine the intentions of the players; then, in the context
of the instantiated BGI with these fixed intentions, we evaluate whether
each $\beta_{i}(t_{i})$ is a best response, just as
in the definition of equilibrium for a standard Bayesian game.

Is this a reasonable notion of equilibrium? As we observed above, in a
BGII, the only possible equilibrium is ``built in'' to the model in the 
form of the intention functions. In particular, the only possible
equilibrium for the instantiation $\tilde{\I}(\beta)$ is $\beta$
itself. Of course, $\beta$ is not necessarily an equilibrium of this
game; however, by quantifying over $\beta$ and considering the
corresponding class of BGIIs (i.e., those obtained as instantiations of
$\tilde{\I}$), we are essentially asking the question: ``Is there a
profile of intentions such that, assuming those intentions are common
knowledge, no player prefers to deviate from their intention?'' If so,
that profile constitutes an equilibrium. This is a natural solution
concept; in fact, as we show in
\fullv{Sections \ref{sec:psy} and
\ref{sec:rdp}, notions of equilibrium proposed by GPS for psychological
games and by K\"oszegi and Rabin \citeyear{KR06} for
reference-dependent preferences,
respectively, are special cases of our definition.}
\shortv{Section \ref{sec:psy}, the notion of equilibrium proposed by
GPS for psychological games is a special case of our definition.}

\begin{example}
{In light of these definitions, Example \ref{exa:sp2} can be viewed as
first defining a BGI $\tilde{\I}$, and then considering two
particular instantiations of it. The equilibrium observations made
then amount to the following: the behaviour rule $\beta_{1} \equiv a$
(i.e., the constant function $a$) is not an equilibrium of
$\tilde{\I}$, but the behaviour rule $\beta_{1}'$ that sends $x$ to
$a$ and $x'$ to $b$ is. (As there is only ever one option for player
$2$'s behaviour rule, namely $\beta_{2} \equiv *$, we can safely neglect
it.)} \qed 
\end{example}

\begin{example}
{Consider again the bravery game of Example \ref{exa:bra}. Under any
particular specification of state space and type spaces, this becomes
a BGI $\tilde{\I}$. It is not difficult to see that each of the
behaviour rules $\beta_{1} \equiv \mathsf{timid}$ and $\beta_{1}'
\equiv \mathsf{bold}$ is an equilibrium of $\tilde{\I}$. \qed} 
\end{example}

\subsection{Existence}

Are equilibria of BGIs guaranteed to exist? Not necessarily. At least
one obstacle to existence lies in the specification of the underlying
type space and the corresponding probability measures: as the
following example shows, certain kinds of belief that are necessary
for best-responses may be implicitly ruled out. 

\begin{example}
{Consider a $2$-player reduced-form BGI $\tilde{\I}$ where
$\Sigma_{1} = \{a,b\}$, $\Sigma_{2} = \{*\}$, $T_{1} = \{x,x'\}$, and
$T_{2} = \{y,y'\}$, and where 
$$p_{1}(x)(\{y\}) = p_{1}(x')(\{y'\}) = p_{2}(y)(\{x\}) = p_{2}(y')(\{x'\}) = 1.$$
Once again we consider a model where player $1$ wishes to surprise her opponent, and so define $u_{1}$ as in Example \ref{exa:sp2}:
$$
u_{1}(\sigma_{1}, *, t) = \left\{ \begin{array}{ll}
1 & \textrm{if $p_{2}(t_{2})(\val{\sigma_{1}}) = 0$}\\
0 & \textrm{otherwise.}
\end{array} \right.
$$
Note that player $1$ is certain that player $2$ knows her type. It
follows that no matter what her intentions are, player $2$ knows them,
and so (by definition of $u_{1}$), player $1$ can always do better by
deviating. In other words, no behaviour rule $\beta_{1}$ is an
equilibrium of $\tilde{\I}(\beta_{1})$ (since it is not a best
response). It follows immediately that $\tilde{\I}$ admits no
equilibria. \qed 
}
\end{example}

This obstacle persists even if we extend our attention to mixed
strategies. More precisely, consider the class of BGIIs where, for each
player $i$, $\Sigma_{i} = \Delta(A_{i})$ for some finite set $A_{i}$
(the set of player $i$'s \emph{pure strategies}), and $u_{i}: \Sigma
\times \Omega \to \mathbb{R}$ satisfies 
$$
u_{i}(\sigma_{i}, \sigma_{-i}, \omega) = \sum_{a_{i} \in A_{i}} \sigma_{i}(a_{i}) u_{i}(a_{i}, \sigma_{-i}, \omega).
$$
In other words, player $i$'s utility for playing $\sigma_{i}$ is just
the expected value of her utility for playing her various pure
strategies with the probabilities given by $\sigma_{i}$. As is
standard, we call elements of $\Sigma_{i}$ \emph{mixed
  strategies}, and the corresponding BGIIs \emph{mixed-strategy BGIIs}.
We can similarly define \emph{mixed-strategy BGIs}.
Note that in this context, since the intention
functions $s_{i}$ map into $\Sigma_{i}$, intended strategies are also
mixed.

The next example shows that, in contrast to the classical setting,
there are mixed-strategy BGIs with finite type spaces that
admit no equilibria. 

\begin{example}
{Consider a $2$-player reduced-form BGI where $\Sigma_{1} =
\Delta(\{a,b\})$, $\Sigma_{2} = \{*\}$, $T_{1} = \{x,x'\}$, and $T_{2}
= \{y,y'\}$, and where 
$$p_{1}(x)(\{y\}) = p_{1}(x')(\{y'\}) = p_{2}(y)(\{x\}) = p_{2}(y')(\{x'\}) = 1.$$
Set
$$
u_{1}(a, *, t) = \left\{ \begin{array}{ll}
1 & \textrm{if $p_{2}(t_{2})(\val{a}) < 1$}\\
0 & \textrm{otherwise}
\end{array} \right.
$$
and
$$
u_{1}(b, *, t) = \left\{ \begin{array}{ll}
1 & \textrm{if $p_{2}(t_{2})(\val{a}) = 1$}\\
0 & \textrm{otherwise,}
\end{array} \right.
$$
and extend to all $\sigma_{1} \in \Delta(\{a,b\})$ by taking expectation:
$$u_{1}(\sigma_{1}, *, t) = \sigma_{1}(a) u_{1}(a,*,t) + \sigma_{1}(b) u_{1}(b, *, t).$$
Note that, following standard conventions, here we identify
the pure strategy $a$ with the degenerate mixed strategy that places
probability $1$ on $a$; likewise for $b$. Thus, for example, the
condition $p_{2}(t_{2})(\val{a}) < 1$ amounts to the following: ``type
$t_{2}$ is not absolutely certain that player $1$ intends to play the
pure strategy $a$'', or equivalently, ``type $t_{2}$ considers it
possible that player $1$ intends to play a mixed strategy that places
positive weight on $b$''. The preferences defined by $u_{1}$ can be
roughly summarized as follows: ``player $1$ prefers to play $a$ in the
event that player $2$ thinks she might place positive weight on $b$,
and prefers to play $b$ if player $2$ is sure that she'll play $a$ for
sure''. 
}

{This game admits no equilibria. To see this, suppose that $\beta_{1}$ were an equilibrium: that is, set player $1$'s intention function equal to $\beta_{1}$, and suppose that $\beta_{1}$ is an equilibrium of the resulting BGII.\footnote{As before, we ignore player $2$'s behaviour since he has no choices to make.} First consider the case where $\beta_{1}(x) \in \Sigma_{1}$ satisfies $\beta_{1}(x)(b) > 0$. Then it follows that $p_{2}(y)(\val{a}) = 0$ (i.e., type $y$ is certain that player $1$ is not playing the pure strategy $a$), and so, since type $x$ is certain that player $2$ is of type $y$, it follows by definition of $u_{1}$ that type $x$'s best response is to play the pure strategy $a$. In particular, $\beta_{1}(x)$ is not a best response, so $\beta_{1}$ cannot constitute an equilibrium. Now consider the case where $\beta_{1}(x)(b) = 0$; in other words, $\beta_{1}(x)$ is the pure strategy $a$. Then we have $p_{2}(y)(\val{a}) = 1$, from which it follows that type $x$'s best response is to play the pure strategy $b$. Thus, once again, $\beta_{1}$ cannot constitute an equilibrium. \qed}
\end{example}

\section{Psychological games} \label{sec:psy}

Psychological games
can be
captured in our framework. A psychological game $\P$ consists of a
finite set of players $N$, together with mixed strategies $\Sigma_{i}$ and
utility functions $v_{i}: \bar{B}_{i} \times \Sigma \to
\mathbb{R}$ for each player $i$,  where $\bar{B}_{i}$ denotes the set
of ``collectively
coherent'' belief hierarchies for player $i$. Somewhat more precisely,
an element $b_{i} \in \bar{B}_{i}$ is an infinite sequence of
probability measures $(b_{i}^{1}, b_{i}^{2}, \ldots)$ where $b_{i}^{1}
\in \Delta(\Sigma_{-i})$ is player $i$'s \emph{first-order beliefs},
$b_{i}^{2}$ is player $i$'s \emph{second-order beliefs} (i.e., roughly
speaking, her beliefs about the beliefs of her opponents), and so on,
such that the beliefs in this sequence satisfy certain technical
conditions (roughly speaking, lower-order beliefs must
agree with the appropriate marginals of higher-order beliefs, and this
agreement must be common knowledge);
see the full paper for the complete definition.

Given a mixed-strategy BGII $\I$ and a type $t_{i} \in T_{i}$, 
we can define the first-order beliefs associated with $t_{i}$ by 
$$\phi_{i}^{1}(t_{i}) = (s_{-i})_{*}(p_{i}(t_{i}));$$
that is, the pushforward of $p_{i}(t_{i})$ from $\Omega$ to
$\Sigma_{-i}$ by $s_{-i}$.
Note that, in our terminology, these are beliefs about
\textit{intended} strategies.
The $k$th-order beliefs associated with
$t_{i}$, denoted $\phi_{i}^{k}(t_{i})$, can be defined inductively in a
similar fashion;
it is then straightforward to show that
the sequence  
$$\phi_{i}(t_{i}) = (\phi_{i}^{1}(t_{i}), \phi_{i}^{2}(t_{i}), \ldots)$$
is collectively coherent, and thus $\phi_{i}: T_{i} \to
\bar{B}_{i}$ (see the full paper).

This correspondence provides a natural notion of
equivalence between psychological games and BGIIs with respect to the
psychological preferences expressed in the former, namely, 
$$
\forall i \in N \ \forall \sigma \in \Sigma \, \forall \omega \in
\Omega (u_{i}(\sigma, \omega) = v_{i}(\phi_{i}(\tau_{i}(\omega)),
\sigma)). 
$$
When a BGII $\I$ satisfies this condition with respect to a psychological game $\P$, we say that $\I$ and $\P$ are \defin{preference-equivalent}.

The notion of preference-equivalence lifts
naturally to BGIs. Observe that the functions $\phi_{i}^{k}$ depend on the profile
of intention functions $s$; being 
explicit about this dependence, we write $\phi_{i}^{k}(t_{i}; s)$ 
rather than $\phi_{i}^{k}(t_{i})$; we then say that $\tilde{\I}$ and
$\P$ are preference-equivalent provided that
$$\forall i \in N \, \forall \sigma \in \Sigma \, \forall
\omega \in \Omega \, \forall s \in \Sigma^T  
  (\tilde{u}_{i}(\sigma, \omega, s) =
  v_{i}(\phi_{i}(\tau_{i}(\omega); s), \sigma)). 
$$

It is easy to see that, given a psychological game $\P$, we can
obtain a preference-equivalent BGI $\tilde{\I}$ simply by
taking the above condition as the \textit{definition} of the
utility functions $\tilde{u}_{i}$. Thus, we have the following:
\begin{proposition} \label{pro:exi}
For every psychological game there exists a
preference-equivalent BGI.
\end{proposition}

Note that even very simple BGIs (i.e., those with very small
type/state spaces) can be preference-equivalent to psychological
games; indeed, it is sufficient for the utility functions $\tilde{u}_{i}$ to
be of the form 
$$\tilde{u}_{i}(\sigma,\omega,s) = f(\phi_{i}(\tau_{i}(\omega);s),\sigma),$$
so that utility depends on states only to the extent that states
encode belief hierarchies.
In particular,
although the utility functions in a psychological game have
uncountable domains (since they apply to all possible belief
hierarchies), a BGI $\tilde{\I}$ can be
preference-equivalent
to a psychological
game $\P$ even if $\tilde{\I}$ has only finitely many states, 
since all that matters is that the
utility functions of $\tilde{\I}$ agree with the utilitiy functions of
$\P$ on the belief hierarchies encoded by the states of $\tilde{\I}$.
Given a psychological game, we can construct a preference-equivalent
BGI with type spaces rich enough that each $\phi_{i}$ is
surjective: in other words, every belief hierarchy is realized
by some type. However, in order to capture \textit{equilibrium}
behaviour, such richness turns out to be superfluous. We now show how
the notion of equilibrium defined by GPS for psychological games can
be recovered as equilibria in our setting.

Given $\sigma \in \Sigma$, let $\chi_{i}(\sigma) \in \bar{B}_{i}$ denote the unique belief hierarchy for player $i$ corresponding to common belief in $\sigma$. A \emph{psychological Nash equilibrium} of $\P$ is a strategy profile $\sigma$ such that, for each player $i$, $\sigma_{i}$
maximizes the function
$$\sigma_{i}' \mapsto v_{i}(\chi_{i}(\sigma),\sigma_{i}',\sigma_{-i}).$$
In particular, to check whether $\sigma$ constitutes a psychological Nash equilibrium, the only relevant belief hierarchies are those corresponding to common belief of $\sigma$. This, in essense, is the reason we do not need rich type spaces in BGIs to detect such equilibria.

\thm \label{thm:psy}
If $\P$ and $\tilde{\I}$ are preference-equivalent, then $\sigma$ is a psychological Nash equilibrium of $\P$ if and only if the profile of (constant) behaviour rules $\beta$ for which $\beta_{i} \equiv \sigma_{i}$ is an equilibrium of $\tilde{\I}$.
\ethm

\prf
When $\beta$ is the profile of behaviour rules described in this
theorem, the corresponding instantiation $\tilde{\I}(\beta)$ has the
property that, for each type $t_{i}$, $\phi_{i}(t_{i}) =
\chi_{i}(\sigma)$. The rest of the proof is essentially just unwinding
definitions; see the full paper for details. 
\eprf

Theorem \ref{thm:psy} shows
that equilibrium analysis in psychological games does not depend on
the full space of belief hierarchies; it can be captured by
particularly simple BGIs. 
%
It
also establishes an equivalence between psychological Nash equilibria and a certain restricted class of equilibria in BGIs; namely, those consisting of constant behaviour rules. This restriction is not surprising in light of the fact that
psychological games do not model strategies as functions of types,
while BGIs do.  Thus, BGIs are not merely recapitulations of the GPS
framework: they are a common generalization of psychological games and
Bayesian games. 


\section{Conclusion}\label{sec:discussion}
We have introduced BGIs, Bayesian games with intentions, which
generalize Bayesian games and psychological games in a natural way.  
We believe that BGIs will prove much easier to deal with than
psychological games, while allowing greater flexibility.

When do equilibria exist?
While Theorem \ref{thm:psy} provides sufficient conditions
for the existence of equilibria in BGIs, they are certainly not 
necessary conditions. We can show, for example, that there are BGIs that 
admit only equilibria in which no behaviour rule is constant. 
Formulating more general conditions sufficient for existence 
is the subject of ongoing work. 

Perhaps the most exciting
prospect for future research lies in leveraging the
distinction between actual and intended
strategies.
As we show in the full paper,
this distinction can be used to implement
K\"oszegi and Rabin's \citeyear{KR06}
model of
reference-dependent preferences;
we believe that it will have other uses as well, and perhaps lead
to new insights into solution concepts.


\section*{Acknowledgements}
We are indebted to Aviad Heifetz for asking the question that first
prompted us to explore the notion of extending Bayesian games to
capture belief-dependent preferences.
Halpern is supported in part by NSF grants
IIS-0911036 and CCF-1214844, by ARO grant W911NF-14-1-0017, 
and by the Multidisciplinary
University Research Initiative (MURI) program administered by the
AFOSR under grant FA9550-12-1-0040.
Pass is supported in part by a Microsoft Research Faculty Fellowship, 
NSF CAREER Award CCF-0746990, AFOSR Award
FA9550-08-1-0197, and BSF Grant 2006317.

\shortv{\bibliographystyle{eptcs}}
\fullv{ \bibliographystyle{chicago}}
\bibliography{z,joe}

\end{document}